\documentclass[ reprint,nofootinbib,amsmath,amssymb,aps]{revtex4-1}
\usepackage{graphicx}
\usepackage{rotating}
\usepackage{dcolumn}
\usepackage{bm}
\usepackage{color}
\usepackage{mathptmx, textcomp}
\usepackage[latin1]{inputenc}
\usepackage{braket}
\usepackage[colorlinks,linkcolor=magenta,anchorcolor=cyan,citecolor=blue]{hyperref}
\usepackage[multiple]{footmisc}

\def\be{\begin{equation}}
\def\ee{\end{equation}}
\def\ba{\begin{eqnarray}}
\def\ea{\end{eqnarray}}

\newcommand{\eq}[1]{(\ref{#1})}

\def\q{\theta} \def\r {\rho}     \def\p {\pi} \def\a {\alpha}    \def\g {\gamma}    \def\l {\lambda}    \def\b {\beta}   \def\pd {\partial}\def\p {\pi}   
\def\Q{\Theta}              \def\grad{\nabla}\def\.{\cdot}
\def\math {\mathcal}
\begin{document}

\title{{Generalized Covariant Entropy Bound in Einstein Gravity with Quadratic Curvature Corrections}}
\author{Hu Zhu$^{1,2}$}
\email{202021140026@mail.bnu.edu.cn}
\author{Jie Jiang$^1$}
\email{Corresponding author. jiejiang@bnu.edu.cn}
\affiliation{${}^1$College of Education for the Future, Beijing Normal University, Zhuhai 519087, China}
\affiliation{${}^2$Department of Physics, Beijing Normal University, Beijing 100875, China}
\date{\today}

\begin{abstract}
We explore the generalized covariant entropy bound in the theory where Einstein gravity is perturbed by quadratic curvature terms, which can be viewed as the first-order quantum correction to Einstein gravity. By replacing the Bekenstein-Hawking entropy with the holographic entanglement entropy of this theory and introducing two reasonable physical assumptions, we demonstrate that the corresponding Generalized Covariant Entropy Bound is satisfied under a first-order approximation of the perturbation from the quadratic curvature terms. Our findings suggest that the entropy bound and the Generalized Second Law of black holes are satisfied in the Einstein gravity under the first-order perturbation from the quadratic curvature corrections, and they also imply that the generalized covariant entropy bound may still hold even after considering the quantum correction of gravity, but in this case, we may need to use holographic entanglement entropy as the formula for gravitational entropy.

\end{abstract}
\maketitle
\section{Introduction}
The discovery of Hawking radiation \cite{Hawking:1974rv} led to the conjecture that black holes can be regarded as thermodynamic systems, possessing entropy fixed at one-fourth of their surface area, i.e., $A/4$. This inspired the formulation of the generalized second law of black hole thermodynamics \cite{Bekenstein:1974ax}. Subsequently, Beckenstein suggested that for the Second Law of Generalized Thermodynamics to hold, the entropy of matter cannot exceed a certain limit, thus proposing the Beckenstein bound \cite{Bekenstein:1980jp}:
\ba
S\leq 2\pi E R/ \hbar
\ea
For spherically symmetric systems in spacetime, to satisfy $E \leq R/2,$ we can substitute this relationship into the above equation, yielding a new constraint:
\begin{equation}\label{40}
S\leq \frac{A}{4}\,.
\end{equation}
Although the arguments put forward for its derivation faced a series of criticisms \cite{Unruh:1982ic,Shimomura:1999xp,Gao:2001ut}, the Beckenstein bound holds in a wide range of physical systems, and it offers profound insights into physics. Early proofs and counterarguments were confined to specific physical systems \cite{Bekenstein:1980jp,Unruh:1982ic,Bekenstein:1982ph,Bekenstein:1984vm,Bekenstein:1993dz,Page:2000uq,Bousso:2003cm,Pelath:1999xt}, without providing a more general proof until 2008 when Casini used quantum field theory to establish the Beckenstein bound in a generic manner in flat spacetime \cite{Casini:2008cr}. As the Beckenstein bound itself applies only to weak gravitational systems, proofs in flat spacetime suffice.

Even before the Beckenstein bound was proven, Bousso was inspired to propose the Covariant Entropy Bound \cite{Bousso:1999xy}, which can be considered as a generalization of the Beckenstein bound to strong gravity systems and is applicable in various curved spacetimes. Specifically, it involves a compact $(D-2)$-dimensional spacelike surface denoted as $B$ with an area represented by $A(B)$. Let $L$ be a null hypersurface generated by null geodesics originating from $B$ and it is not terminated until a caustic point is reached. Assuming that the expansion of the null congruence is nonpositive, we have
\[S_L\leq \frac{A(B)}{4},\]
where $S_L$ represents the entropy passing through  $L$.

It is worth noting that the conjecture mentioned above stipulates that the light sheet $L$ must terminate after reaching a focal point. Consequently, Flanagan and others \cite{Flanagan:1999jp} extended this entropy bound by allowing the light sheet to terminate at another $(D-2)$-dimensional spatial surface $B'$. This leads to the modified entropy bound known as the generalized covariant entropy bound or generalized Bousso bound:
\ba\begin{aligned}\label{ebound}
S_L\leq \frac{1}{4}\left|A(B')-A(B)\right|\,,
\end{aligned}\ea
where $A(B')$ represents the area of the spatial surface $B'$. It is important to note that this bound has been proven within the framework of Einstein gravity, subject to certain physical assumptions \cite{Flanagan:1999jp,Bousso:2003kb,Strominger:2003br}.

However, these proofs and the form of the Covariant Entropy Bound are based on classical Einstein gravity. On the one hand, considering that classical gravity theory will not be the ultimate theory of gravity and, in the context of effective field theories with quantum corrections and string theory, Einstein gravity will include effective corrections due to high-curvature terms \cite{A13,A14,A15,A16}. On the other hand, both the Covariant Entropy Bound and the Bekenstein Bound take quantum effects into account, rendering the conjecture semi-classical in nature when defined in a classical gravity background. To provide a more consistent understanding, the first step is to extend the Covariant Entropy Bound to more general gravity theories, and this has been discussed in a recent article \cite{Matsuda:2020yvl}. In their article, the authors replaced the area term with a general black hole entropy formula, resulting in a new Covariant Entropy Bound. However, its effectiveness has only been demonstrated in specific cases, such as spherically symmetric configurations in $f(R)$ gravity and Gauss-Bonnet gravity. Considering that high-curvature terms can only act as small corrections to Einstein gravity, especially in the context of effective field theories, where the order of curvature terms is proportional to the order of perturbations, we focus primarily on the Covariant Entropy Bound in first-order quantum corrections to Einstein gravity (Einstein gravity with second-order curvature corrections). This consideration is particularly relevant in scenarios where high-curvature terms are treated as perturbations, which, in turn, is consistent with the effective field theory framework for quantum corrections. We aim to demonstrate the effectiveness of the extended Covariant Entropy Bound in the context of first-order quantum corrections to Einstein gravity (i.e., the Einstein gravity with second-order curvature corrections).

The structure of the paper is organized as follows. In Sec. \ref{sec2}, we provide a concise overview of quadratic gravity. Sec. \ref{sec3} introduces the Generalized Entropy Bound within the framework of Einstein gravity perturbed by quadratic curvature corrections. We outline the underlying physical assumptions and critical aspects necessary to establish this bound. In Sec. \ref{sec4},  we present the proof of the Generalized Entropy Bound in a theory where second-order curvature terms perturb Einstein gravity. We illustrate that holographic entanglement entropy rigorously complies with the entropy bound, even under the influence of first-order perturbations in second-order curvature. Finally, Sec. \ref{sec5} concludes the paper and offers a discussion of the findings.

\section{Quadratic gravity}\label{sec2}

In this paper, we consider the Einstein gravity with second-order curvature correction. The action of this theory in $D$-dimensional spacetime is given by
\ba\begin{aligned}
I=\frac{1}{16\p}\int d^Dx\sqrt{g}\left(R+ a R^2+ b R_{ab}R^{ab}+c \math{L}_\text{GB}+\math{L}_\text{mat}\right),
\end{aligned}\ea
in which $\math{L}_\text{mat}$ is the Lagrangian density of the matter fields, $g_{ab}$ is the Minkowski metric of the spacetime, and
\ba\begin{aligned}
\math{L}_\text{GB}=R^2-4R_{ab}R^{ab}+R_{abcd}R^{abcd}
\end{aligned}\ea
is the Gauss-Bonnet term. Here, three coupling constants $a, b, c$ are independent, and they are all very small. It is worth noting that in both $f(R)$ gravity and Einstein-Gauss-Bonnet gravity, the Generalized Covariant Entropy Bound has already been established \cite{Matsuda:2020yvl, Zhang:2022yvv}. Therefore, we only need to demonstrate that the $R_{ab}R^{ab}$ term, when considered as a standalone modification, also holds.
\begin{equation}\label{59}
\mathcal{L}=\frac{1}{16\pi}\left(R+\l R_{a b} R^{a b}+\math{L}_\text{mat}\right)\,.
\end{equation}

Based on the Lagrangian, we can derive the field equations as
\ba\label{eom}
G_{a b}+\l H_{a b}=8 \pi T_{a b}\,.
\ea
Here, $G_{a b}$ represents the Einstein tensor, and $H_{a b}$ expands as follows:
\begin{equation}\label{EOM}\begin{aligned}
H_{a b}&=\nabla^{c}\nabla_{c} R_{a b}-\nabla_{a} \nabla_{b} R+2 R_{a c b d} R^{c d}\\
&+\frac{1}{2} g_{a b} \left(\nabla^{c}\nabla_{c} R-R_{c d} R^{c d}\right)\,,
\end{aligned}\end{equation}
in which $T_{ab}$ is the stress-energy tensor  of the matter fields.

\section{Generalized covariant entropy bound}\label{sec3}

First we clarify the definition of the covariant entropy bound: we use two compact $(D-2)$-dimensional spatial surface $B_0$ and $B_1$ to represent the boundaries of the initial and final physical systems, there is a null hypersurface $L$ generalized by null geodesics, which starts at $B_0$ and ends at $B_1$. The tangent vector field of the null geodesics calls $k^a=(\pd/\pd u)^a$, $u$ is an affine parameter, choose it so that the spatial surfaces $B_0$ and $B_1$ are given by $u=0$ and $u=1$. We require that the expansion $\q$ associated with $k^a$ is non-positive throughout $L$. Then, select a coordinate system on $L$, denotes $(u, x)$, $x=\{x^1,\cdots x^{D-2}\}$ is the coordinate of the cross-section, and every geodesic is determined by a constant $x$. The covariant entropy bound in those setups is that the entropy $S_L$ passing $L$ should satisfy
\ba\begin{aligned}\label{eb}
S_L\leq |S_\text{grav}(B_0)-S_\text{grav}(B_1)|\,,
\end{aligned}\ea
in which $S_\text{grav}(B)$ is an type of entropy on the cross-section $B$, and we choose it so that the entropy bound is satisfied.

We further introduce the Gaussian null coordinate system $\{z, u, x\}$, the line element can be expressed as
\ba\begin{aligned}\label{lineelement}
ds^2(\l)=2(dz+z^2\a du+z \b_i dx^i)du+\g_{ij}dx^i dx^j\,,
\end{aligned}\ea
where the null hypersurface $L$ is given by $z=0$, and $\a$, $\b_i$ and $\g_{ij}$ are the function of $u,z,x,\l$. The index $i, j, k, l$ denotes the spacelike coordinates.

To discuss the covariant entropy bound in a gravitational system, we first need to understand how the entropy mentioned in the above inequality is defined in the context of that gravitational system. Using the Noether charge method of Iyer and Wald \cite{A17,A18}, the Wald entropy is given by
\ba\begin{aligned}
S_W=-2\p\int_s d^{D-2}x\sqrt{\g}\frac{\pd \math{L}}{\pd R_{abcd}}\hat{\bm{\epsilon}}_{ab}\hat{\bm{\epsilon}}_{cd}\,,
\end{aligned}\ea
where $s$ is a cross-section of event horizon, $\g_{ab}$ is the induced metric on $s$, and $\hat{\bm{\epsilon}}_{ab}$ is the binormal to $s$ and it gives the corrected first law of black holes. However, as discussed in Refs. \cite{A25,A26,A27,A28,Bhattacharjee:2015yaa,Wall:2015raa}, the Wald entropy of the quadratic gravity does not obey the linearized second law and we need to focus on the gravitational entropy as
\ba\label{JMS}
S_\text{grav}=\frac{1}{4}\int_s d^{D-2}x\sqrt{\g} \r
\ea
with
\ba\begin{aligned}
\r=1+2\l R_{zu}-\frac{\l}{2}K\bar{K}\,,
\end{aligned}\ea
in which
\ba\begin{aligned}
K_{ij}=\frac{1}{2}\pd_u \g_{ij}\,, \quad\quad \bar{K}_{ij}=\frac{1}{2}\pd_z \g_{ij}
\end{aligned}\ea
are the extrinsic curvature of the null vectors $(\pd/\pd u)^a$ and $(\pd/\pd z)^a$ separately. This is also the holographic entanglement entropy in quadratic gravity \cite{Dong:2013qoa}. Next, we will use the this entropy mentioned to prove the Generalized Covariant Entropy Bound.

The generalized expansion $\Q$ of the entropy is defined by
\ba\begin{aligned}
\frac{d S_\text{grav}}{d u}=\frac{1}{4}\int_B d^{D-2}x \sqrt{\g} \Q\,,
\end{aligned}\ea
which gives the change of the gravitational entropy per unit area. Using the field equation \eq{eom}, it is straightforward to obtain:
\ba\begin{aligned}\label{dQdu}
\frac{d\Q}{d u}=-8\p \math{T}+\math{F}\,,
\end{aligned}\ea
in which we denote
\ba\begin{aligned}
\math{T}&=T_{ab}k^a k^b\,,\\
\math{F}&=G_{ab}k^a k^b+\l H_{ab}k^ak^b+k^a \grad_a \Q\,.
\end{aligned}\ea
This can be seem as the Raychaudhuri equation in the quadratic gravity.

In the thermodynamic limit, the entropy passing through the light sheet $L$ is given by
\ba\begin{aligned}
S_L=\int_{L}d^{D-2}x du\sqrt{\g} s
\end{aligned}\ea
with the entropy density $s=-k_a s^a$, in which $s^a$ is the entropy flux field of the matter field. 

As suggested in Ref. \cite{Matsuda:2020yvl}, we introduce two following assumptions in the quadratic gravity,
\ba\begin{aligned}\label{assumption}
&\text{(i)}\quad \pd_u s(x,u)\leq 2\p \math{T}(x,u)\,,\\
&\text{(ii)}\quad s(x, 0)\leq -\frac{1}{4}\Q(x, 0)\,
\end{aligned}\ea
on $L$. The assumptions mentioned above were also introduced in Einstein gravity \cite{Bousso:2003kb,Strominger:2003br}. Assumption (i) was introduced to ensure that the rate of change of entropy flux is less than or equal to the energy flux, which can also be seen as a result of the Bekenstein bound's version \cite{Matsuda:2020yvl}. Assumption (ii) is simply an initial choice of the hypersurface to ensure that the entropy bound is valid at the beginning of $L$.

Using the above setups, it is not hard to get
\ba\begin{aligned}
&s(x, u)=s(x, 0)+\int_{0}^{u} du\pd_u s(x, u)\\
&\leq s(x, 0)-\frac{1}{4}\Q(x, \l)+\frac{1}{4}\Q(x, 0)+\frac{1}{4}\int_{0}^{u}d\tilde{u} \math{F}(x, \tilde{u})\\
&\leq-\frac{1}{4}\Q(x, \l)+\frac{1}{4}\int_{0}^{u}d\tilde{u} \math{F}(x, \tilde{u})\,.\\
\end{aligned}\ea
Integration of the above identity over $L$ gives
\ba\begin{aligned}\label{ineq11}
S_L\leq& S_\text{grav}(B_0)-S_\text{grav}(B_1)\\
&+\frac{1}{4}\int_{0}^{1}du \int_{0}^{u}d\tilde{u} \int d^{D-2}x\sqrt{\g(u)}\math{F}(\tilde{u},x)\,.
\end{aligned}\ea
For the Einstein gravity, this is just the Raychaudhuri equation and it is easy to see $\math{F}(\tilde{u},x)\leq 0$, which implies that the Generalized Covariant Entropy Bound is satisfied in Einstein gravity. For the other cases, if we can show $\math{F}(\tilde{u},x)\leq 0$, the entropy bound will be hold. That is to see, the key point to finish the entropy bound is to judge the sign of $\math{F}$.

\section{Proof of the entropy bound with quadratic curvature corrections}\label{sec4}

From the perspective of quantum corrections and string theory, Einstein gravity is subject to corrections from high-curvature terms. Therefore, in the following discussion, we treat the quadratic term in the action as a small perturbation to Einstein gravity, assuming that $\l$ is a small quantity that describes the perturbation from high-curvature terms. Next, we will explore the covariant entropy bound in the first-order perturbative approximation of $\l$, specifically calculating the sign of $\math{F}$ at this order.

From the field equation \eq{EOM} and Eq. \eq{dQdu}, we can easily get
\ba\begin{aligned}\label{dF}
\math{F}&=-K_{b}^a K_a^b+\frac{\l}{4}\math{F}_2\,.
\end{aligned}\ea
with
\begin{equation}\label{110}\begin{aligned}
\math{F}_2=&\left[\nabla_{c}\nabla^{c}R_{a b}-\nabla_{a}\nabla_{b}R+2R_{a c b d}R^{c d}\right] k^{a}k^{b}\\
&+2\frac{\partial}{\partial u} \left[ \frac{1}{\sqrt{\gamma}} \frac{\partial}{\partial u} \sqrt{\g} \left(R_{z u}-\frac{1}{2} K \bar{K}\right) \right]\,.
\end{aligned}\end{equation}

Using the line element \eq{lineelement}, it is not hard to get
\begin{equation}\label{expG}
\begin{aligned}
\Gamma_{i j}^k & =\hat{\Gamma}_{i j}^k, \quad \Gamma_{u i}^j=K_i^j, \quad \Gamma_{z i}^j=\bar{K}_i^j, \quad \Gamma_{u z}^1=\frac{1}{2} \beta^i\,, \\
\Gamma_{i j}^u & =-\bar{K}_{i j}, \Gamma_{u i}^u=-\frac{1}{2} \beta_i, \Gamma_{i j}^z=-K_{i j}, \Gamma_{z i}^z=\frac{1}{2} \beta_i\,,
\end{aligned}
\end{equation}
and
\ba\begin{aligned}\label{expR}
&R_{ij}^{kl}=\hat{R}_{ij}^{kl}-4K_{[i}^{[k}\bar{K}_{j]}^{l]}\,,\quad R^{zj}_{ui}=-\pd_u K_i^j-K_{i}^kK^{j}_k\,,\\
&R_{ui}^{jk}=-2D^{[j}K_i^{k]}+K_i^{[j}\b^{k]}\,,\quad R^{zi}_{jk}=-2D_{[j}K^i_{k]}+K^i_{[j}\b_{k]}
\end{aligned}\ea
at the light sheet $L$. Here $\hat{R}_{ijkl}$ is the spatial curvature tensor on $B$.

Using the above results, we can obtain
\begin{equation}\label{111}\begin{aligned}
\math{F}_2 =& k^{a}k^{b}\nabla_{c}\nabla^{c}R_{a b}-R,_{u u}+2R_{u c u d}R^{c d}\\
&+2 R_{z u},_{u u}+2K,_{u} R_{z u}-\left(K \bar{K}\right),_{u u}+{o}(K)\\
=&k^{a}k^{b}\nabla_{c}\nabla^{c}R_{a b}+2R_{u c u d}R^{c d}-R_{i}^{i},_{u u}\\
&+2K,_{u} R_{z u}-\left(K \bar{K}\right),_{u u}+\math{O}\left(K\right)\,,
\end{aligned}\end{equation}
where $O(K)$ denotes the first and higher-order term of $K_a^b$, such as $K_a^b R_b^a$ and $K$,  but it does not include spatial and temporal derivatives of $K_a^b$. In the last step, we used the identity $R=2R_{zu}+R^{i}_{i}$ from the GN line element \eq{lineelement}. Using the GN coordinates, the first term becomes
\begin{equation}\label{113}
\begin{aligned}
k^{a}k^{b}\nabla_{c}\nabla^{c}R_{a b}=&\nabla_{c}\nabla^{c}R_{u u}-2\nabla_{c}k^{a}  \nabla^{c}k^{b}  R_{a b}\\
&-2k^{a}\nabla_{c}\nabla^{c}k^{b} R_{a b}-4k^{a}\nabla_{c}k^{b} \nabla^{c}R_{a b} \\
=&2 R_{uu},_{zu}+A
\end{aligned}
\end{equation}
in which the specific expression for $A$ will be provided later. Then, we have
\begin{equation}\label{F2exp}
\begin{aligned}
\math{F}_2=&A+2R_{u c u d}R^{c d}+2 R_{uu},_{zu}-R_{i}^{i},_{u u}+2K,_{u} R_{z u}\\
&-\left(K \bar{K}\right),_{u u}+o(K) \\
=&A+2R_{u c u d}R^{c d}+2 R_{uzu}^{z},_{zu}+2R_{uiu}^{i},_{zu}-2 R_{z i u}^{i},_{u u}\\
&-R_{j i j}^{i},_{u u}+2K,_{u} R_{z u}-\left(K \bar{K}\right),_{u u}+\math{O}\left(K\right)
\end{aligned}
\end{equation}
Next, we calculate each part separately. Transform the covariant derivatives into partial derivatives and express curvature tensors using Christoffel symbols:
\begin{equation}\label{115}\begin{aligned}
\nabla_{a} w_{b}&=\partial_{a} w_{b}+\Gamma_{ab}^{c}w_{c}\,,\\
R_{\mu v \sigma}^\rho&=\Gamma_{\mu \sigma,v}^\rho-\Gamma_{v \sigma,\mu}^\rho+\Gamma_{\sigma \mu}^\lambda \Gamma_{v \lambda}^\rho-\Gamma_{\sigma v}^\lambda \Gamma_{\mu \lambda}^\rho\,.
\end{aligned}\end{equation}
Finally, expand the Christoffel symbols in the chosen coordinate system. It is worthy noting that the Christoffel symbol expressions in Eqs. \eq{expG} and \eq{expR} are only valid on the light sheet $L$. If there are derivatives with respect to $z$, separate calculations are required. Additionally, since the final result is an integration over $L$ with spatial components on a closed surface, some terms that can be expressed as complete spatial derivative terms can be omitted, specifically:
\begin{equation}\label{115}
\int_B e \sqrt{\gamma} f=\int_B e F^{i}_{,i}=0
\end{equation}
Where $e$ represents the coordinate element of $B$. Also, terms like $K_{,i} f$ can be transformed into $(K f)_{,i}-K f_{,i}$, where the first part can be omitted, and the second part contains first-order terms involving $K$, i.e., we can treat
\ba
K_{,i} f=\math{O}(K)\,.
\ea
It should be noted that when performing these reductions, the expressions need to be multiplied by $\sqrt{\gamma}$ to ensure that the resulting terms are complete spatial derivatives.

For the first term of Eq. \eq{F2exp}, we have
\begin{equation}\label{116}
\begin{aligned}
A=&-4\alpha K,_{u} \\
&-\gamma^{i j}\bar{K}_{i j}K,_{u u} \\
&+\frac{3}{2}\beta^{i},_{u}\beta_{i},_{u}+\frac{3}{2}\beta^{i}\beta_{i}K,_{u}+\beta^{i}\beta_{i},_{uu} \\
&+\beta^{i}K,_{u i}-\beta_{i},^{i} K,_{u}-2\beta^{i}K_{i}^{j},_{j u}-\beta^{i}K_{i}^{k},_{u}\gamma^{j l}\gamma_{j l},_{k}\\
&+\beta^{i}K^{k l},_{u} \gamma_{k l},_{i}+\frac{1}{2}\beta^{i}K,_{u}\gamma^{j k}\gamma_{j k},_{i}\,.
\end{aligned}
\end{equation}
For the second term of Eq. \eq{F2exp}, we have
\begin{equation}\label{117}
\begin{aligned}
2R_{u c u d}R^{c d}=&4\alpha K,_{u} \\
&+4K_{i j},_{u}\bar{K}^{i j},_{u} \\
&-\frac{1}{2}\beta^{i}\beta_{i}K,_{u}-\beta^{i},_{u}\beta_{i},_{u}+K_{i j},_{u} \beta^{i}\beta^{j} \\
&-2\beta^{i},^{j} K_{i j},_{u}-\beta^{k} K_{i j},_{u} \gamma^{i j},_{k} \\
&-2K_{i l},_{u} \hat{R}^{i l}\,.
\end{aligned}
\end{equation}
For the third term of Eq. \eq{F2exp}, we have
\begin{equation}\label{118}
2R_{u c u d}R^{c d}=-\frac{1}{2}\gamma_{i j},_{u u} \beta^{i} \beta^{j}-\left(\beta_{i},_{u} \beta^{i} \right),_{u}\,.
\end{equation}
For the fourth and fifth terms of Eq. \eq{F2exp}, we have
\begin{equation}\label{119}
\begin{aligned}
2R_{uiu}^{i},_{zu}+2 R_{z i u}^{i},_{u u}=&-4\alpha K,_{u} \\
&+2\left(K^{i j} \bar{K}_{i j}\right),_{uu}-2\left(K^{i j} K_{i j}\right),_{zu} \\
&-\frac{1}{2}\beta^{j} \gamma^{i k},_{u u}\gamma_{i k},_{j}-\frac{1}{2}\beta^{j} \gamma^{i k}\gamma_{i k},_{u u j}\,.
\end{aligned}
\end{equation}
For the sixth term of Eq. \eq{F2exp}, we have
\begin{equation}\label{120}
\begin{aligned}
-R_{j i j}^{i},_{u u}=&2\left(K \bar{K}\right),_{uu}-2\left(K^{i j} \bar{K}_{i j}\right),_{uu} \\
&-\hat{R},_{u u}\,.
\end{aligned}
\end{equation}
For the last term of Eq. \eq{F2exp}, we have
\begin{equation}\label{120}
\begin{aligned}
2K,_{u} R_{z u}=&4\alpha K,_{u} \\
&-2K,_{u}\bar{K},_{u} \\
&-\beta^{i}\beta_{i}K,_{u} \\
&-\frac{1}{2}\beta^{k}K,_{u}\gamma^{i j}\gamma_{i j},_{k}+\beta_{i},_{j}K,_{u}\gamma^{i j}\,.
\end{aligned}
\end{equation}
To keep the notation simple, we omit the $\math{O}\left(K\right)$  terms  in the calculations presented above and in the following.

Now, the goal is to cancel out these terms. The $\alpha$ terms in the first, second, fourth, fifth, and seventh lines of the first term can be seen to cancel directly. Next is the $K\bar{K}$ term, which can also be canceled. Due to the simplicity of the calculation, this is not listed here. The double $\beta$ term is also straightforward and results in
\ba
-\frac{1}{2}\beta^{i}_{,u}\beta_{i,u}\,,
\ea
which is negative.

The remaining task is to show that the first line of the single $\beta$ term and the terms from the third line sum to zero, and the second line of the single $\beta$ term combined with the second line of the second term can be canceled out:
\begin{equation}\label{121}
-2K_{i l},_{u} \hat{R}^{i l}-\hat{R},_{u u}=-\gamma_{i l},_{u u} \hat{R}^{i l}-\hat{R},_{u u}=\gamma^{i l},_{u u} \hat{R}_{i l}-\hat{R},_{u u}\,,
\end{equation}
where the first step is to express $K$ using $\g$. The second step is because for the first-order derivative of $\g$, we have
\begin{equation}\label{122}
\gamma_{i j},_{u}=-\gamma_{i k}\gamma_{j l}\gamma^{k l},_{u}\,.
\end{equation}
Notice that $\gamma_{,u}\propto K$, so the two-order derivative of $\gamma_{,uu}$ has a similar transformation. For the same reason, we can further obtain:
\begin{equation}\label{124}
\gamma^{i l},_{u u} \hat{R}_{i l}-\hat{R},_{u u}=(\gamma^{i l} \hat{R}_{i l}),_{u u}-\gamma^{i l} \hat{R}_{i l},_{u u}-\hat{R},_{u u}=-\gamma^{i l} \hat{R}_{i l},_{u u}\,.
\end{equation}
Moreover, considering $\partial_x (dy)^a = 0$, we have
\begin{equation}\label{125}
-\gamma^{i l} \hat{R}_{i l},_{u u}=-\gamma^{a b} \hat{R}_{a b},_{u u}\,.
\end{equation}
Since $\hat{R}$ is fully determined by $\gamma$, its second-order derivative can only expressed by $\gamma$, which is actually a variation of the metric $\gamma_{ab}$ in the $(D-2)$-dimensional space, i.e.,  $\gamma^{ab} \delta \hat{R}_{ab}$. Using the identity
\begin{equation}\label{126}
g^{a c} \delta R_{a c} =\nabla^a\left(\nabla^b \delta g_{a b}-g^{b c} \nabla_a \delta g_{b c}\right)\,,
\end{equation}
we can get
\begin{equation}\label{127}\begin{aligned}
-\gamma^{a c} \hat{R}_{a c},_{u u} &=-\tilde{\nabla}^a\left(\tilde{\nabla}^b \gamma_{a b},_{u u}-\gamma^{b c} \tilde{\nabla}_a \left(\gamma_{b c},_{u u}\right)\right)\\
&=-\tilde{\nabla}^a v_{a}\,,
\end{aligned}\end{equation}
where $\tilde{\nabla}$ is the covariant derivative of the $(D-2)$-dimensional surface $B$. Considering the integral of this equation on the surface:
\begin{equation}\label{128}
\int_B e \sqrt{\gamma} \tilde{\nabla}^a v_a=\int_B \bm\varepsilon \tilde{\nabla}^a v_a\,,
\end{equation}
where $\bm\varepsilon$ is the adapted element of the $(D-2)$-dimensional surface $B$. According to Gaussian theorem, it is obviously zero.

Finally, only the single $\beta$ term is left. After some obvious reduction, we can obtain:
\begin{equation}\label{129}
\begin{aligned}
&\beta^{i}K,_{u i}-2\beta^{i}K_{i}^{j},_{j u}-\beta^{i}K_{i}^{k},_{u}\gamma^{j l}\gamma_{j l},_{k}+\beta^{i}K^{k l},_{u} \gamma_{k l},_{i} \\
&-2\beta^{i},^{j} K_{i j} \\
&-\frac{1}{2}\beta^{j} \gamma^{i k}\gamma_{i k},_{u u j}\,,
\end{aligned}
\end{equation}
where the first line corresponds to the first term, the second line corresponds to the second term, and the third line corresponds to terms four and five. All other terms are obviously canceled. Now, we will evaluate that the first and fourth terms of first line, combined with the third line, sum up to zero, and second and third term of first line together with the second line's terms can reduce to
\begin{equation}\label{130}
\begin{aligned}
&\beta^{i}K,_{u i}+\beta^{i}K^{k l},_{u} \gamma_{k l},_{i}-\frac{1}{2}\beta^{j} \gamma^{i k}\gamma_{i k},_{u u j}=0 \\
&-2\beta^{i}K_{i}^{j},_{j u}-\beta^{i}K_{i}^{k},_{u}\gamma^{j l}\gamma_{j l},_{k}-2\beta^{i},^{j} K_{i j}\,.
\end{aligned}
\end{equation}
Expand the first term:
\begin{equation}\label{131}
\begin{aligned}
\beta^{i}K,_{u i}&=\frac{1}{2}\beta^{i}\left(\gamma^{j k} \gamma_{j k},_{u}\right),_{u i}\\
&=\frac{1}{2}\beta^{i}\gamma^{j k},_{i}\gamma_{j k},_{u u}+\frac{1}{2}\beta^{i}\gamma^{j k}\gamma_{j k},_{u u i} \\
&=\frac{1}{2}\beta^{i}\gamma_{j k},_{i}\gamma^{j k},_{u u}+\frac{1}{2}\beta^{i}\gamma^{j k}\gamma_{j k},_{u u i}\\
&=-\beta^{i}\gamma_{j k},_{i}K^{j k},_{u}+\frac{1}{2}\beta^{i}\gamma^{j k}\gamma_{j k},_{u u i}\,.
\end{aligned}
\end{equation}
This precisely cancels out with the remaining terms. The second term needs to be multiplied by $\sqrt{\gamma}$, and we will demonstrate that it is a total space derivative
\begin{equation}\label{132}
\begin{aligned}
&-2\beta^{i}K_{i}^{j},_{j u}\sqrt{\gamma}-\beta^{i}K_{i}^{k},_{u}\gamma^{j l}\gamma_{j l},_{k}\sqrt{\gamma}-2\beta^{i},^{j} K_{i j}\sqrt{\gamma} \\
&=-2\beta^{i}K_{i}^{j},_{j u}\sqrt{\gamma}-2\beta^{i}K_{i}^{k},_{u}\sqrt{\gamma},_{k}-2\beta^{i},^{j} K_{i j}\sqrt{\gamma} \\
&=-2\left(\beta^{i}K_{i}^{j},_{u}\sqrt{\gamma} \right),_{j}+2\beta^{i},_{k} K_{i}^{k},_{u}\sqrt{\gamma}-2\beta^{i},^{j} K_{i j}\sqrt{\gamma}\\
&=-2\left(\beta^{i}K_{i}^{j},_{u}\sqrt{\gamma} \right),_{j}\,,
\end{aligned}
\end{equation}
in which the first step is due to
\ba
\frac{1}{2}\gamma^{i j}\gamma_{i j},_{k}=\frac{1}{\sqrt{\gamma}}\pd_k\sqrt{\gamma}\,.
\ea
The final step is also due to the fact that the first derivative of $\gamma$ with respect to $u$ cancels out automatically, so the indices can be ignored.

Finally, we can find that, apart from the terms of order $K$, the correction introduces only the quadratic $\beta$ term, $-(1/2)\beta^{i}_{,u} \beta_{i,u}$. Therefore, we have
\begin{equation}\label{78}
\math{F}=-\frac{1}{4} K_a^bK_{b}^a-\frac{1}{8}\l \gamma^{i j} \beta_{i},_{u} \beta_{j},_{u}+\l \math{O}\left(K\right)\,.
\end{equation}

From the previous discussion, we know that we are considering perturbations of high curvature terms on Einstein gravity, where $\lambda$ acts as a small parameter. Therefore, we only need to determine the sign of $\math{F}$ in this perturbed scenario. Since the second-order curvature terms are just first-order quantum corrections, we are here considering only the first-order approximation in $\lambda$, i.e., neglecting $O(\lambda^2)$.

In the case of a perturbation expansion, the sign of $F$ depends solely on the leading term in the entire expansion. Assume that
\ba
K_a^bK_{b}^a = \lambda^{2s}.
\ea
Then, using the Cauchy-Schwarz inequality, we have
\ba\begin{aligned}
|\math{O}(K)|= |X_a^b K_a^b|\leq \sqrt{X_a^bX_b^a}\sqrt{K_a^bK_b^a}=\math{O}(\lambda^{s})\,.
\end{aligned}\ea
If $s<1/2$, the leading term is given by
\ba\begin{aligned}
\math{F}\simeq -\frac{1}{4}K_a^bK_{b}^a
\end{aligned}\ea
which is negative. If $1/2\leq s < 1$, considering $\l \math{O}(K) = \math{O}(\l^{s+1})$ and $2s<s+1$, we have
the leading term is given by
\ba\begin{aligned}
\math{F}\simeq -\frac{1}{4} K_a^bK_{b}^a-\frac{1}{8}\l \gamma^{i j} \beta_{i},_{u} \beta_{j},_{u} < 0\,.
\end{aligned}\ea
Finally, if $s>1$, under the first-order approximation of $\l$, i.e., neglecting $O(\lambda^2)$, we have
\ba\begin{aligned}
\math{F}\simeq-\frac{1}{8}\l \gamma^{i j} \beta_{i},_{u} \beta_{j},_{u}+O(\lambda^2) \leq 0\,.
\end{aligned}\ea
In summary, our results indicate that, under the first-order perturbation of high curvature terms, the leading term of $\math{F}$ is always negative. This means that the covariant entropy bound holds for Einstein gravity under the first-order perturbation of second-order curvature terms.

\section{Conclusion and discussion}\label{sec5}

In the context of effective field theories with quantum corrections, Einstein gravity incorporates effective corrections arising from high-curvature terms \cite{A13,A14,A15,A16}. The order of curvature terms is proportional to the order of perturbations. To account for quantum effects, we primarily focus on the Covariant Entropy Bound in Einstein gravity, perturbed by quadratic curvature corrections. Introducing a small parameter $\lambda$ to characterize these perturbations, we equivalently consider the Covariant Entropy Bound under a first-order approximation of $\lambda$. Upon examining the linearized second law of black holes in quadratic gravity, we naturally propose the entropy bound in this theory by replacing the Bekenstein-Hawking entropy with holographic entropy.

By introducing two reasonable physical assumptions, we demonstrate that the validity of the Generalized Covariant Entropy Bound is equivalent to $\math{F}\leq0$. Consequently, we establish that the leading term of $\math{F}$ is always positive under the first-order approximation of the quadratic curvature perturbation, i.e., neglecting $O(\lambda^2)$. In other words, the Generalized Covariant Entropy Bound is satisfied under the first-order quantum corrections. Our findings suggest that the entropy bound and the Generalized Second Law of black holes are satisfied in Einstein gravity under the first-order perturbation from quadratic curvature corrections. This implies that the Generalized Covariant Entropy Bound may remain valid even after considering quantum corrections to gravity and we may need to use holographic entanglement entropy as the formula for gravitational entropy.

\section*{Acknowledgement}
This paper is supported by the National Natural Science Foundation of China with Grant No. 12205014, the
Guangdong Basic and Applied Research Foundation with Grant No. 2021A1515110913, and the Talents Introduction
Foundation of Beijing Normal University with Grant No. 310432102.

\end{document}